\documentclass[aps,preprintnumbers,floatfix,article,amsmath,amssymb,floatfix,10pt,prd,superscriptaddress,nofootinbib]{revtex4-2}
\usepackage{bm}
\usepackage{amsfonts}
\usepackage{latexsym}
\usepackage[latin1]{inputenc}
\usepackage{graphicx}
\usepackage{amsmath}
\usepackage{palatino}
\usepackage{mathpazo}
\usepackage{textcomp}
\usepackage{float}
\usepackage{booktabs}
\usepackage{dcolumn}
\usepackage{ragged2e}
\usepackage{hyperref}
\hypersetup{colorlinks,citecolor=purple}
\hypersetup{colorlinks=true,linkcolor=red,filecolor=magenta,    urlcolor=blue}
\usepackage{amsmath}
\usepackage{xcolor}
\usepackage{epsfig}
\usepackage{caption}
\usepackage{subcaption}
\usepackage{commath}
\def\jnl@style{\it}
\def\aaref@jnl#1{{\jnl@style#1}}

\def\aaref@jnl#1{{\jnl@style#1}}

\def\aj{\aaref@jnl{AJ}}                   
\def\apj{\aaref@jnl{ApJ}}                 
\def\apjl{\aaref@jnl{ApJ}}                
\def\apjs{\aaref@jnl{ApJS}}               
\def\apss{\aaref@jnl{Ap\&SS}}             
\def\aap{\aaref@jnl{A\&A}}                
\def\aapr{\aaref@jnl{A\&A~Rev.}}          
\def\aaps{\aaref@jnl{A\&AS}}              
\def\mnras{\aaref@jnl{Mon.~Not.~Roy.~Astron.~Soc.}}             
\def\prd{\aaref@jnl{Phys.~Rev.~D}}        
\def\prc{\aaref@jnl{Phys.~Rev.~C}}  
\def\prl{\aaref@jnl{Phys.~Rev.~Lett.}}    
\def\qjras{\aaref@jnl{QJRAS}}             
\def\skytel{\aaref@jnl{S\&T}}             
\def\ssr{\aaref@jnl{Space~Sci.~Rev.}}     
\def\zap{\aaref@jnl{ZAp}}                 
\def\nat{\aaref@jnl{Nature}}              
\def\aplett{\aaref@jnl{Astrophys.~Lett.}} 
\def\apspr{\aaref@jnl{Astrophys.~Space~Phys.~Res.}} 
\def\physrep{\aaref@jnl{Phys.~Rep.}}      
\def\physscr{\aaref@jnl{Phys.~Scr}}       
\def\commat{\aaref@jnl{Comm.~Math.~Phys.}}              
\def\science{\aaref@jnl{Science}}               
\def\cqg{\aaref@jnl{Classical Quant.~Grav.}}            
\def\jpcs{\aaref@jnl{JPCS}}                                     
\def\ijmpd{\aaref@jnl{Int.~J.~Mod.~Phys.~D}}                    
\def\grg{\aaref@jnl{Gen.~Relat.~Gravit.}}               
\def\rpp{\aaref@jnl{Rep.~Prog.~Phys.}}          
\def\npa{\aaref@jnl{Nucl.~Phys.~A}}        
\def\lrr{\aaref@jnl{Living Rev.~Rel.}}                   
\def\jcap{\aaref@jnl{J.~Cosmology Astropart.~Phys.}}    
\def\rmp{\aaref@jnl{Rev.~Mod.~Phys.}}   
\def\epjc{\aaref@jnl{Eur.~Phys.~J.~C}} 
\def\plb{\aaref@jnl{~Phy.~Lett.~B}} 
\def\mpla{\aaref@jnl{Mod.~Phy.~Lett.~A}} 
\def\arxiv{\aaref@jnl{arxiv.org}}

\allowdisplaybreaks[1]

\begin{document}
\title{\bf Dynamical features of $f(T,B)$ cosmology}

\author{S. A. Kadam}
\email{k.siddheshwar47@gmail.com}
\affiliation{Department of Mathematics,
Birla Institute of Technology and Science-Pilani, Hyderabad Campus,
Hyderabad-500078, India.}

\author{B. Mishra}
\email{bivu@hyderabad.bits-pilani.ac.in}
\affiliation{Department of Mathematics,
Birla Institute of Technology and Science-Pilani, Hyderabad Campus,
Hyderabad-500078, India.}

\author{S.K. Tripathy}
\email{tripathy\_sunil@rediffmail.com}
\affiliation{Department of Physics, Indira Gandhi Institute of Technology, Sarang, Dhenkanal, Odisha-759146, India.}

\begin{abstract}
\textbf{Abstract}: In this paper, we have explored the field equations of $f(T,B)$ gravity and determined the dynamical parameters with the hyperbolic function of Hubble parameter. The accelerating behaviour has been observed and the behaviour of equation of state parameter indicates $\Lambda CDM$ model at late time. The role of model parameters in assessing the accelerating behaviour has been emphasised. Interestingly the term containing $\beta$, the coefficient of boundary term, in the model parameter vanishes during the simplification. The scalar perturbation has been presented to show the stability of the model.
\end{abstract}
\keywords{}
\maketitle
\textbf{Keywords}: $f(T,B)$ gravity, Energy conditions, Cosmographic parameters, Stability analysis.

\maketitle
\section{Introduction}

Several cosmological observations such as, Supernovae Ia \cite{Perlmutter99,Riess98}, cosmic microwave background radiation \cite{Hinshaw07}, large scale structure \cite{{Tegmark04}} have shown the path breaking result of accelerated expansion of the Universe. At the present stage, Universe has already been entered into the accelerating phase. In order to explain the reason, researchers have predicted the presence of dark matter and dark energy in the Universe, whose form is still to be established. The present cosmological observation claims that $68.3$ percent of the Universe is filled with dark energy \cite{Aghanim20}. Theoretically to describe the dark energy, the vacuum energy or the cosmological constant model is most appealing since this model agrees with the observational predictions. However, it has certain difficulties to reconcile the small observational value of dark energy density that comes from quantum field theories\cite{Capozziello11}. This is known as the cosmological constant problem. Therefore, it has been believed that modification of General Relativity (GR) would provide significant contribution to explain the late time cosmic acceleration \cite{Bahamonde15,Bahamonde17}. Post supernovae era, we have seen several modifications of GR and one important modification has been performed by introducing torsion scalar $T$ in place of usual curvature term $R$ in the action. This modification is known as the teleparallel gravity, or the teleparallel equivalent of GR \cite{Weitzenbock23,Hehl76}. In this approach, the gravitational interaction has been described by force equivalent to the Lorentz force equation of electrodynamics. In GR and its extensions, the metric components are obtained from Levi-Civita connection, which is curvature-ful whereas in teleparallel gravity, the tetrad components are derived from Weitzenb$\ddot{o}$ck connection, that is torsion-ful. Further the teleparallel gravity has been extended to $f(T)$ gravity \cite{Ferraro07,Fiorini11}, which can be built similar to $f(R)$ extension of GR  \cite{Felice10,Odinstov11}. To mention here, the equations of motion in $f(T)$ gravity are of second-order whereas in $f(R)$ gravity it is of fourth order. So, the torsion based gravity has been preferred. More torsion based modified theories of gravity are suggested, such as $f(T,\mathcal{T})$, where $\mathcal{T}$ is the trace of energy-momentum tensor \cite{Saez-Gomez16}, $f(T,T_{G})$ where $T_{G}$ is torsion scalar equivalent of Gauss Bonnet term $G$ \cite{{Kofinas14,Chattopadhyay14}}, $f(T,\phi)$ gravity \cite{Espinozal21} where $\phi$ is scalar field, $f(T,B)$ gravity where $B$ is the boundary term which relates torsion scalar and Ricci scalar together as $R=-T+B$ \cite{Bahamonde15,Bahamonde17}. Amongst these modified teleparallel gravity theories, $f(T,B)$ gravity ensures its viability on various scales ranging from solar system test \cite{Farrugia20}, Noether symmetry approach \cite{Bahamonde17} and confronting with the observational data \cite{Escamilla-Rivera20}. Our study would explore the possibility of obtaining an accelerating theoretical model within the framework of $f(T,B)$ gravity with some known form of the Hubble function. More specifically we shall analyse the cosmological parameters to be obtained and will compare with the prescribed value of cosmological observations. For more information on the teleparallel gravity from theory to cosmology, someone can refer to the interesting review \cite{Bahamonde21}.\\
  
We shall discuss some of the recent works on the teleparallel  family of gravitational theories. To extract the perturbed equations of motion, the scalar perturbation in $f(T)$ gravity has been examined \cite{Chen11}. The late time cosmic acceleration model can be obtained in $f(T)$ gravity \cite{Myrzakulov11}. The validity of generalized second law of gravitational thermodynamics in the framework of $f(T)$ gravity has been shown in Ref. \cite{Karamia12}. The stability of the gravitational theory in FLRW space-time has also been examined in Ref. \cite{Farrugia16}. In presence of collisional matter, the phantom divide line crossing can be realized in $f(T)$ gravity in form of power law, exponential law and logarithmic functions \cite{Zubair16}. The reconstruction in scalar-tensor theory has been analysed, and the inhomogeneous equation of state parameter was estimated using a particular Hubble parameter to describe early and late-time epoch in Ref. \cite{Said17}. Otalora \cite{Otalora17} have examined the covariant $f(T)$ theories on its admitting of G$\ddot{o}$del-type solutions. In Lorentz-covariant formalism, the cosmological solutions of $f(T)$ gravity can be feasible and with Bianchi identities, the compatible solutions can be shown \cite{Golovnev20}. The gravitational dynamics of braneworld models governed by teleparallel $f(T)$ gravity has also been studied \cite{Moreira21a}.\\

Teleparallel equivalent of general relativity (GR) has an associated Lagrangian which is equivalent to general relativity up to a boundary term \cite{Pereira01,Maluf13}. So, the dynamical equations are same as that of GR though sourced by different gravitational actions. The modification in the theory arises from the boundary term between GR and TEGR. In GR, these boundary terms arises naturally since the second order derivatives appear in its Lagrangian. With the boundary term $B$, the $f(T,B)$ gravity has been formulated \cite{Bahamonde15}.The boundary term is the source of the fourth orders derivative. In $f(T,B)$ gravity the second and fourth order derivative contributions to the field equations contribute independently to the gravitational action. The cosmological reconstruction has been performed in the FLRW background and with the cosmological reconstruction the power law, de-Sitter and $\Lambda$CDM models can be realised \cite{Bahamonde18}.  In Ref.\cite{Capozziello20},  in presence of matter,  the exact field equations of $f(T,B)$ gravity can be obtained and in a low energy limit, it can be linearised. This will further allow to get the gravitation waves. The dynamical variables are compatible with the late time cosmic observations and are stable \cite{Franco20}. On the basis of jerk parameter, the cosmological importance of this gravity has been worked upon \cite{Zubaira20}. In Ref. \cite{Caruana20}, the cosmological bounce can be realised, thereby  possible occurrence of singularity can be avoided. At the same time, it has been revealed that  torsion and boundary term affects the thick brane scenario \cite{Moreira21b}. Paliathanasis \cite{Paliathansis21} obtained the solution of the gravitational equations and investigated the existence of quantum corrections for the gravitational field equations. Hence, we are here motivated to study the late time cosmic acceleration issue in the $f(T,B)$ gravity with some form of the Hubble function. The paper is organised as: in Sec. \ref{sec:background} we have presented the field equations of $f(T,B)$ gravity and its dynamical parameters. In Sec. \ref{sec:Hyperbolic_scale_factor}, we have incorporated the Hubble parameter, whose scale factor describes some hyperbolic function. The late time cosmic acceleration behaviour has been presented through the equation of state (EoS) parameter. Also, we have derived the energy conditions and have shown its behaviour in the context of extended teleparallel gravity. In Sec. \ref{sec:Cosmography Parameters}, the geometrical parameters are analysed and the stability of the model in the scalar perturbation approach has been discussed. Finally the results and conclusion are given in Sec. \ref{sec:conclusion}. 

\section{\texorpdfstring{$f(T,B)$}{} Gravity Field Equations}\label{sec:background}

In this section, we shall present a brief discussions on the $f(T,B)$ gravity followed by deriving the expressions for the dynamical parameters. The tetrad fields or the vierbein $e^A_{\mu}$ are the dynamical variables. At each point of the space time manifold, these dynamical variables will form an orthonormal basis for the tangent space. So, the tetrad and its inverse preserves the orthogonality relations as, 

\begin{eqnarray}
e^{A}_{~~\mu}\ \ e_{B}^{~~\mu}&=&\delta^{B}_{A}\,, \nonumber\\
e^{A}_{~~\mu} \ \ e^{~~\nu}_{A}&=&\delta^{\nu}_{\mu}, \label{eq:1} 
\end{eqnarray}

where $e^{\mu}_B$ be the inverse of $e^A_{\mu}$. The metric tensor $g_{\mu\nu}$ can be obtained from the tetrad fields as,

\begin{eqnarray}
g_{\mu \nu}&=& e^{A}_{~~\mu} e^{B}_{~~\nu}\eta_{AB}, \nonumber \\
\eta_{AB}&=& e^{~~\mu}_{A} e^{~~\nu}_{B} g_{\mu \nu}, \label{eq:2}
\end{eqnarray} 
where $\eta_{AB}$ represents the Minkowski metric. The teleparallel connection $\Gamma_{\mu \nu}^{\sigma}$ can be expressed with respect to the tetrads and spin connection $\omega_{n\mu}^{m}$ as \cite{Weitzenbock23}, 
\begin{equation}
     e_{A}^{\sigma}\left(\partial_{\mu} e^{A}_{\nu} + \omega^{A}_{B \mu} e^{B}_{\nu}\right)=\hat{\Gamma}^{\sigma}_{\nu\mu} \,\label{eq:3}
\end{equation}
and the non-zero torsion tensor can be written as, 
\begin{equation}
\hat{\Gamma}^{\lambda}_{\nu \mu}-
\hat{\Gamma}^{\lambda}_{\mu\nu}=T^{\lambda}_{\mu\nu}\,,\label{eq:4}
\end{equation}
The property of torsion scalar $T$ allows us to express it as the product of superpotential and torsion tensor as, 
\begin{equation}
S_{\sigma}^{\ \ \mu\nu}T^{\sigma}_{\ \ \mu\nu}=T\,,\label{eq:5}
\end{equation}
where the superpotential term, 
\begin{equation}
(K^{\ \ \mu\nu}_{\sigma}-\delta^{\mu}_{\sigma}T^{\nu}+\delta^{\nu}_{\sigma}T^{\mu})=2S_{\sigma}^{\ \ \mu\nu}\,,\label{eq:6}
\end{equation}

where $K^{\mu\nu}_{\ \ \sigma}$ be the contorsion tensor and
\begin{equation}
(T^{\sigma}_{ \ \ \mu\nu}-T^{ \ \ \sigma}_{\nu\mu}+T^{\ \ \sigma}_{\mu\ \ \nu})=2K^{\ \ \sigma}_{\mu\ \ \nu}\,,\label{eq:7}
\end{equation}
From Eq. \eqref{eq:5}, the torsion scalar can be obtained in term of Hubble parameter as, 
\begin{equation}
T=6H^2 \,.\label{eq:8}
\end{equation}
where the spin connection represents the degrees of freedom associated with the local Lorentz transformation invariance of the theory and are zero in the so-called Weitzenb\"{o}ck gauge of the connection. As mentioned before, the Ricci scalar and the boundary term differs by a boundary term through the relation $R=-T+B=-T+ \frac{2}{e}\partial_{\mu}(eT^{\mu})$, where $e$ is the determinant of the tetrad. Hence, the action of Teleparallel Equivalent of GR reproduces the same field equations as in GR. Considering that the function $f(T)$ depends on the boundary term $B$, the action of $f(T,B)$ gravity has been given as \cite{Bahamonde15},

\begin{equation}
S_{f(T,B)}=\frac{1}{\kappa^2} \int d^4x e f(T,B)+\mathcal{L}_m\,,\label{eq:9}
\end{equation}
with $\kappa^2= 8\pi G$ and $\mathcal{L}_m$ be the matter Lagrangian. Subsequently, varying the action in Eq.~\eqref{eq:9} with respect to the tetrad fields, the $f(T,B)$ gravity field equations can be derived as \cite{Bahamonde15,Bahamonde17}
\begin{eqnarray}
& & e_A{}^{\mu} \square f_B -  e_A {}^{\nu} \nabla ^{\mu} \nabla_{\nu} f_B +
	\frac{1}{2} B f_B e_A{}^{\mu} - \left(\partial _{\nu}f_B + \partial
	_{\nu}f_{T} \right)S_A{}^{\mu\nu}  \nonumber \\
& & -\frac{1}{e} f_{T}\partial _{\nu} (eS_A{}^{\mu\nu})
	+ f_{T} T^{B}{}_{\nu A}S_{B}{}^{\nu\mu}- f_T \omega ^B{}_{A\nu}
	S_B{}^{\nu\mu} -\frac{1}{2}  f E_{A}{}^{\mu} =  \kappa ^2  \Theta _A{}^{\mu} \,,\label{eq:10}
\end{eqnarray}
where $f_T$ and $f_B$ respectively denote partial derivatives with respect to $T$ and $B$. Also, $\Theta_{m}^{\lambda}$ and $\nabla_{\sigma}$ are respectively denote the energy momentum tensor and Levi-Civita covariant derivative with respect to the Levi-Civita connection. Consider a tetrad for the flat FLRW metric as,
\begin{equation}\label{eq:11}
    e_{\mu}^{m}=(1,a(t),a(t),a(t))\,,
\end{equation}
which satisfies the Weitzenb\"{o}ck gauge for $f(T,B)$ gravity. We wish to study the cosmological aspects of $f(T,B)$ gravity in an isotropic and homogeneous background, 
\begin{equation}\label{eq:12}
    ds^2=-dt^2+a^2(t)(dx^2+dy^2+dz^2)\,,
\end{equation}

Also, we consider that the Universe is filled with the matter in the form of perfect fluid. Then the field equations of $f(T,B)$ gravity \eqref{eq:10} for the tetrad field \eqref{eq:11} and metric \eqref{eq:12} can be derived as, 
\begin{eqnarray}
    -3H^2(3f_B+2f_T)+3H \dot{f_B}-3\dot{H}f_B+\frac{1}{2}f(T,B)=\kappa^2\rho\,,  \label{eq:13}\\
    -3H^2(3f_B+2f_T)-\dot{H}(3f_B+2f_T)-2H\dot{f_T}+\ddot{f_B}+\frac{1}{2}f(T,B)=-\kappa^2 p\,. \label{eq:14}
\end{eqnarray}

Here, an overdot denotes the derivative with respect to time $t$. $p$ and $\rho$ respectively denote the effective pressure and effective energy density. The logarithmic  function form is able to analyse the equation of state for dark energy in teleparallel modified gravity formalism\cite{Bamba11} and also describe stable de sitter solution in the dynamical system analysis for certain parametric range\cite{Zhang11}. To analyse the fate of the universe, we probe the behaviour of EoS parameter with a linearly coupled boundary term in the modified gravity theory,  which may specifically be expressed in the form of $f(T,B)$ as $f(T,B)=\alpha T log(\frac{T}{T_{0}})+\beta B$. The coefficients $\alpha$, $\beta$ and $T_{0}$ are constants. The expression between the Hubble parameter and scale factor, $H=\frac{\dot{a}}{a}$ simplifies $T=6H^2$ and $B=6\left(\dot{H}+3H^2 \right)$. Now, the expression for effective EoS parameter which will enable us to analyse the dark energy phase can be obtained from Eqs. \eqref{eq:13}-\eqref{eq:14} as,

\begin{equation}
\omega=\frac{p}{\rho}=-1+\frac{3H\dot{f}_B+2\dot{H}f_T+2H\dot{f}_T}{-9H^2f_B-6H^2f_T+3H\dot{f}_B-3\dot{H}f_B+\frac{1}{2}f(T,B)} \label{eq:15}
\end{equation}

To solve the system, an appropriate relationship between the matter term or the metric potentials to be considered. Since in FLRW space time isotropic with equal metric potentials, we prefer to consider a known form of the scale factor in the subsequent section.

\section{The Cosmological Model}\label{sec:Hyperbolic_scale_factor}
\begin{figure}[H]
\centering
\includegraphics[width=80mm]{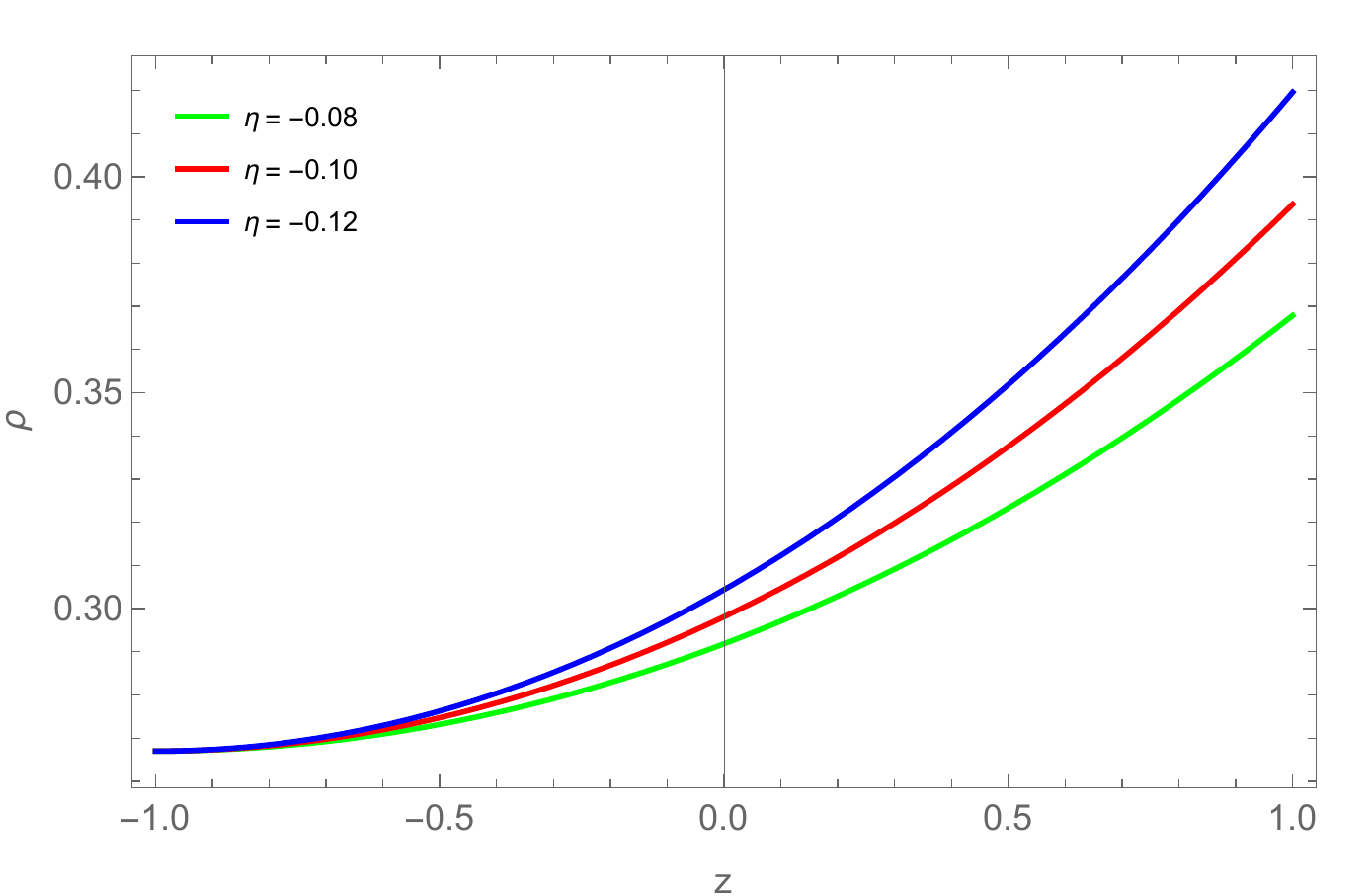}
\includegraphics[width=80mm]{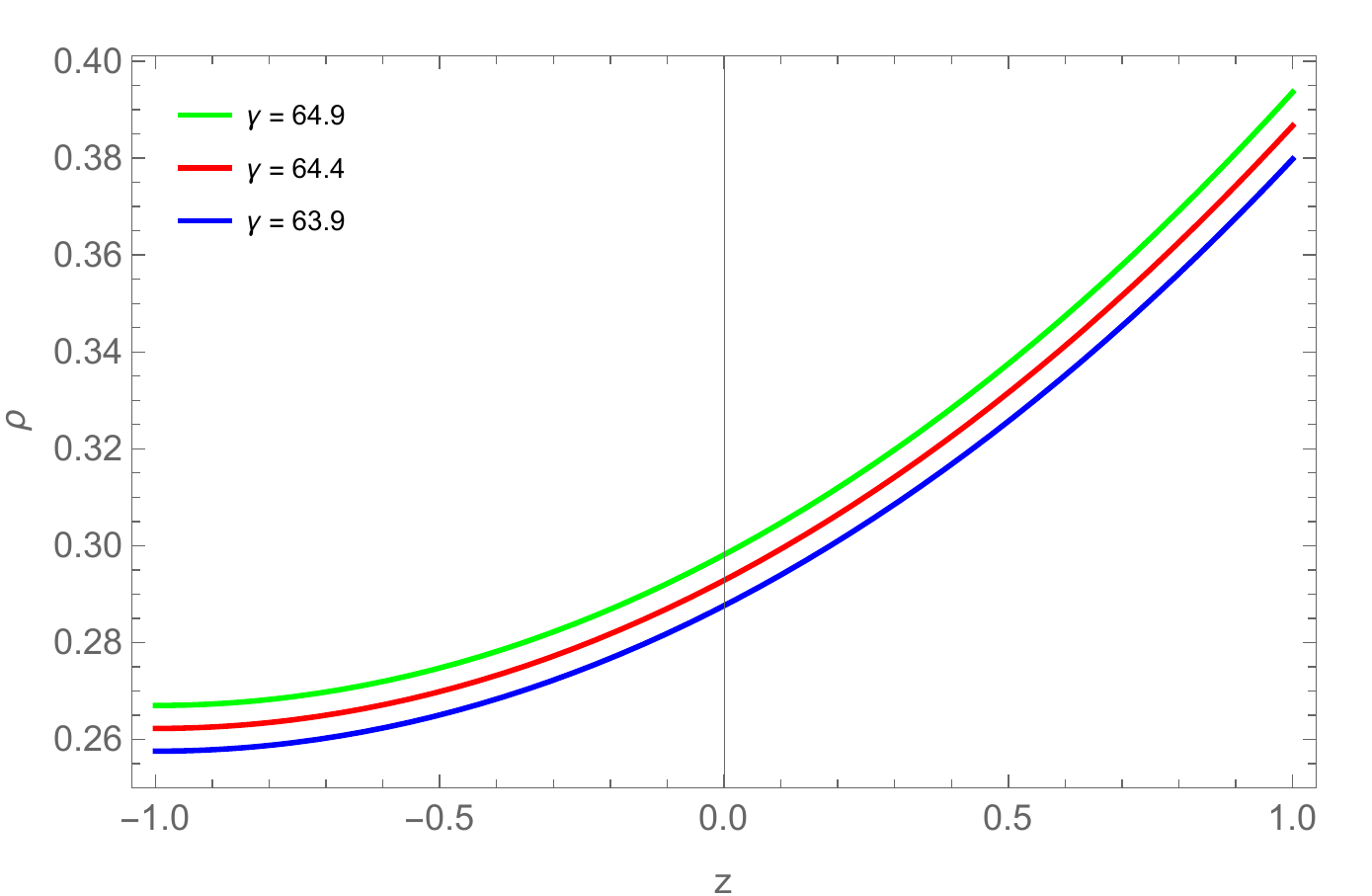}
\caption{Energy density in redshift for varying $\eta$(left panel) and $\gamma$(right panel). Other parameters, $\alpha=-0.05$, $T_{0}=105 \pi$.} 
\label{Fig1}
\end{figure}

In order to study the background cosmology, the field equations \eqref{eq:13}-\eqref{eq:14} and the EoS parameter \eqref{eq:15} need to be analysed. All these expressions contain terms those can be expressed in the form of Hubble parameter. We consider the hyperbolic form of the Hubble parameter as, $H(t)=\gamma tanh(\gamma t)$, where $\gamma$ is a constant . Also the classical redshift-scale factor relation provides, $1+z=\frac{1}{a}$ and subsequently the Hubble parameter can be parameterized in redshift function as, $H(z)=\gamma \left[1-(1+z)^{2}\eta\right]^{\frac{1}{2}}$, where $\eta$ is a constant. Hence \eqref{eq:13}-\eqref{eq:15} can be expressed in terms of redshift as,
\begin{eqnarray}    
p(z)&=&\alpha  \gamma ^2 \left(6-\left(\eta  (z+1)^2-3\right) \log \left(-\frac{6 \gamma ^2 \left(\eta  (z+1)^2-1\right)}{T_{0}}\right)\right), \label{eq:16}\\
\rho(z)&=&3 \alpha  \gamma ^2 \left(\eta  (z+1)^2-1\right) \left(\log \left(-\frac{6 \gamma ^2 \left(\eta  (z+1)^2-1\right)}{T_{0} }\right)+2\right) , \label{eq:17}  \\
\omega(z)&=&\frac{6-\left(\eta  (z+1)^2-3\right) \log \left(-\frac{6 \gamma ^2 \left(\eta  (z+1)^2-1\right)}{T_{0} }\right)}{3 \left(\eta  (z+1)^2-1\right) \left(\log \left(-\frac{6 \gamma ^2 \left(\eta  (z+1)^2-1\right)}{T_{0}}\right)+2\right)}. \label{eq:18}
\end{eqnarray}
Since the expressions are complicated to analyse the behaviour of the parameters, we adopted here the graphical representation of the parameters. Also, the behaviours of the dynamical parameters depend on the parametric values of the scale factor and $f(T,B)$. In the accelerating model, the pressure and energy density respectively should behave negative and positive and subsequently the EoS parameter to be negative at present and late time. Accordingly the parametric values of the scale factor are chosen. Since the scale factor contains two parameters, we consider some representative values of both the parameters in the model. It is interesting to note here that the terms containing the model parameter $\beta$ vanishes identically during the simplification. The energy density remains positive throughout the evolution with varying $\eta$ FIG. \ref{Fig1} (left panel) and varying $\gamma$ FIG. \ref{Fig1} (right panel). In both cases, the energy density is found to decrease slowly from early to late time. The uniform behaviour of the varying $\gamma$ values is maintained through out the evolution, the curve with higher $\gamma$ remaining at the top. On the other hand, the energy density curves with lower values of $\eta$ remains at the top at an early epoch and possibly at the bottom at late times. However all the energy density curves for different $\eta$ values appear to merge at late times of evolution. The curve shows slight shifting at an early time. Since the Universe was denser in the early era and it gets expanding throughout its evolution results in lowering the energy density.  

\begin{figure}[H]
\centering
\includegraphics[width=80mm]{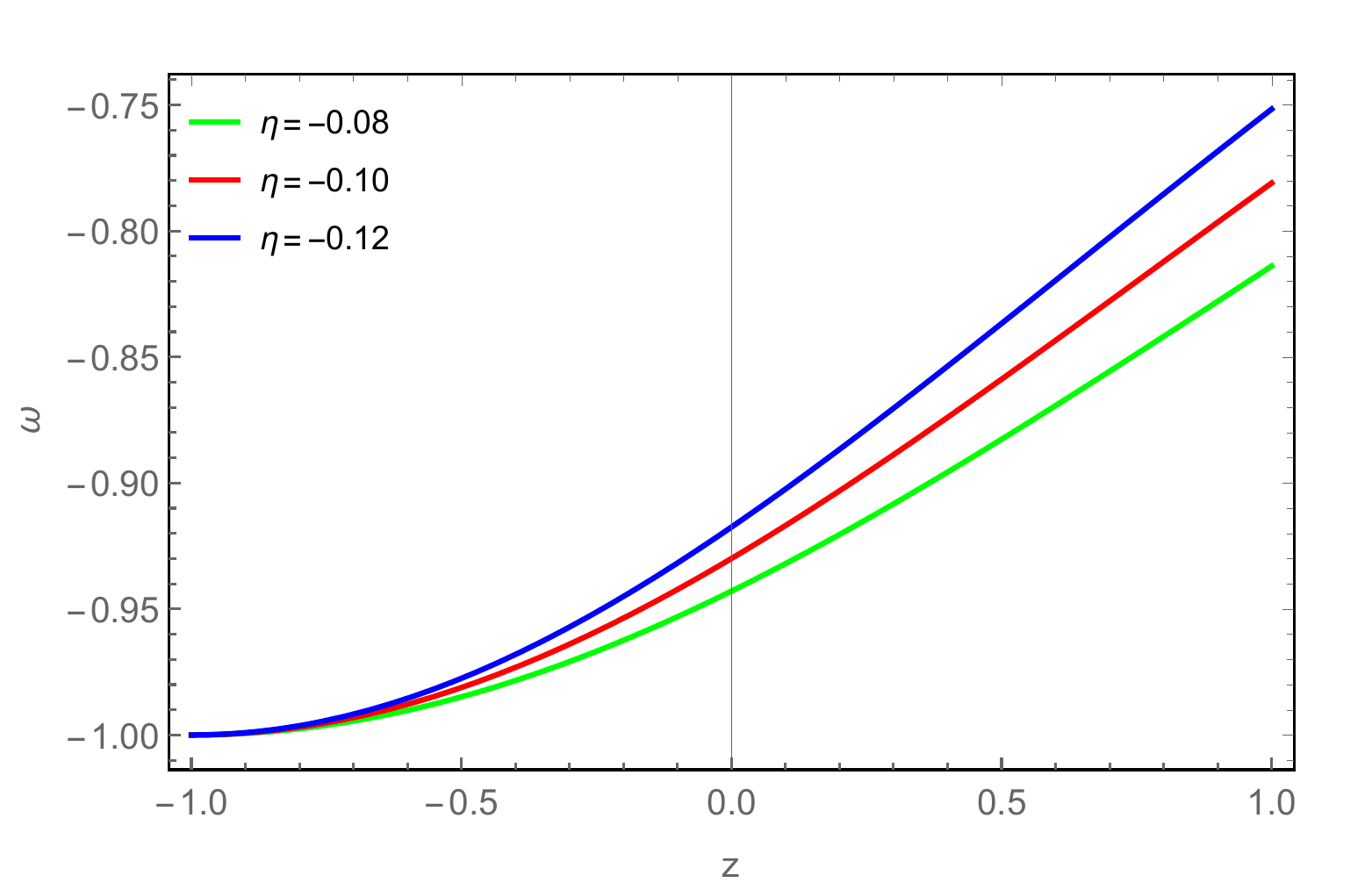}
\includegraphics[width=80mm]{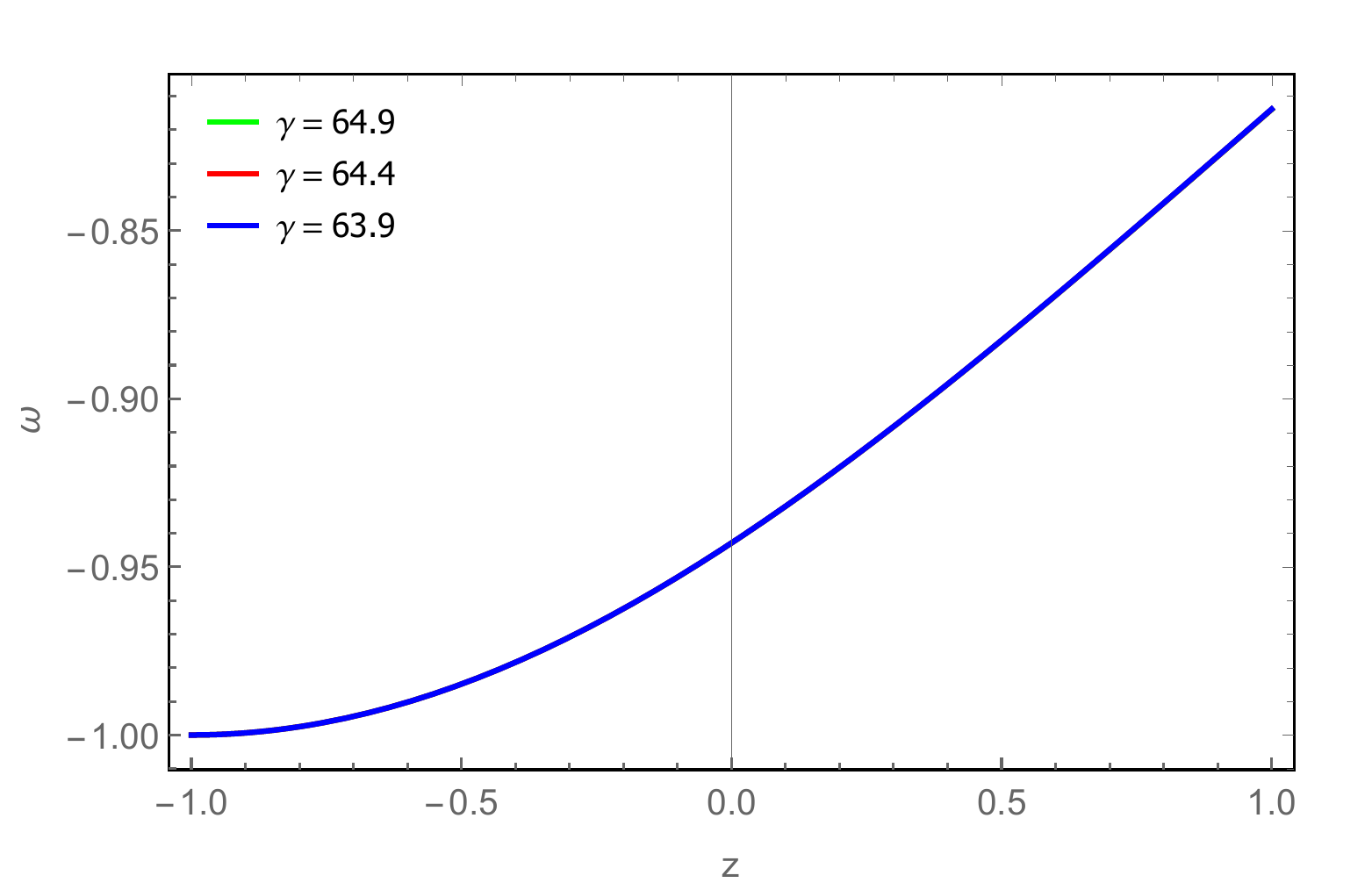}
\caption{EoS parameter in redshift for varying $\eta$(left panel) and $\gamma$(right panel).Other parameters, $\alpha=-0.05$, $T_{0}=105 \pi$.} 
\label{Fig2}
\end{figure}

To probe the dynamics of the universe, the graphical behaviour of EoS parameter would be assessed. It is worthy to mention here some of the recent cosmological observations on the present value of EoS parameter: Supernovae Cosmology Project, $\omega_0=-1.035 ^{+0.055}_{-0.059}$ \cite{Amanullah10}; WMAP+CMB, $\omega_0=-1.073 ^{+0.090}_{-0.089}$ \cite{Hinshaw13}; Planck Collaboration $\omega_0:-0.957\pm0.080$ (Planck+SNe+BAO), $\omega_0:-0.76\pm0.20$(Planck+BAO/RSD+WL), Planck 2018, $\omega_0=-1.03\pm 0.03$ \cite{Aghanim20}. In addition to this, to cater the $H_o$ tension issue, the following values of EoS parameter have been suggested: $\omega_0: -1.29^{+0.15}_{-0.12}$ \cite{Valentino16}, $\omega_0\approx-1.3$ \cite{Vagnozzi20}. In the present work, we have obtained the EoS parameter using a hyperbolic Hubble function and plotted the graphs by varying the parameters $\eta$ and $\gamma$. For the representative values of the parameter $\eta$, it has been observed that all the curves decrease from low negative value to high negative value and almost merge together at late times to the concordant $\Lambda$CDM value of $-1$ thereby showing the $\Lambda$CDM behaviour at late phase. The present value of the EoS parameter noted respectively for $\eta=-0.08,-0.10,-0.12$ as $\omega_0=-0.942,-0.929,-0.916$ [FIG. \ref{Fig2} (left panel)]. At the same time, there is no significant difference noticed in the evolution behaviour of EoS parameter with a variation in $\gamma$. All the curves are merged together and approach to $-1$ at late time leading to $\Lambda$CDM behaviour with the present value noted as $\omega_0=-0.943$ [FIG. \ref{Fig2} (right panel)]. So it has been observed that this model remains at the $\Lambda$CDM phase at late time of evolution irrespective of their behaviour at early time. Also FIG.\ref{Fig2} indicates that the the EoS parameter might have experienced a zero value at an early phase of the evolution of Universe. Hence  study of logarithmic form of the shape function in modified teleparallel gravity theories may help us to describe the matter dominated era.

Since the EoS parameter suggests the $\Lambda$CDM model, we wish to verify the behaviour of energy conditions of the model. The energy conditions are: Null Energy Condition (NEC): $\rho+p \ge 0$, Weak Energy Condition (WEC): $\rho+p \ge 0$, $\rho \ge 0$, Strong Energy Condition (SEC): $\rho+3p \ge 0$ and Dominate Energy Condition (DEC): $\rho-p \ge 0$. Energy conditions can have a significant impact on cosmic evolution. One important result is that the limits imposed by the energy conditions can be linked to the acceleration and deceleration phenomena of cosmic fluid and the possible occurrence of Big Rip singularities \cite{Capozziello18}. In an expanding Universe, the NEC confirms the decreasing behavior of energy density, as well as NEC and SEC are important assumptions of the Penrose singularity theorem \cite{Rubakov14,Sharifa16}. DEC plays an important role in the proof of positive mass theorem \cite{{Schoen81}}. The expressions for the energy conditions can be obtained from Eqs. \eqref{eq:16}-\eqref{eq:17} as,

\begin{eqnarray}\label{eq:19} 
\rho(z)+p(z)&=&2 \alpha  \gamma ^2 \eta  (z+1)^2 \left(\log \left(-\frac{6 \gamma ^2 \left(\eta  (z+1)^2-1\right)}{{\color{blue}T_{0}}}\right)+3\right), \nonumber \\
\rho(z)+3p(z)&=&6 \alpha  \gamma ^2 \left(\log \left(-\frac{6 \gamma ^2 \left(\eta  (z+1)^2-1\right)}{{\color{blue}T_{0} }}\right)+\eta  (z+1)^2+2\right) \nonumber \\
\rho(z)-p(z)&=&2 \alpha  \gamma ^2 \left(\left(2 \eta  (z+1)^2-3\right) \log \left(-\frac{6 \gamma ^2 \left(\eta  (z+1)^2-1\right)}{{\color{blue}T_{0} }}\right)+3 \eta  (z+1)^2-6\right).  
\end{eqnarray} 

\begin{figure}[H]
\centering
\includegraphics[width=80mm]{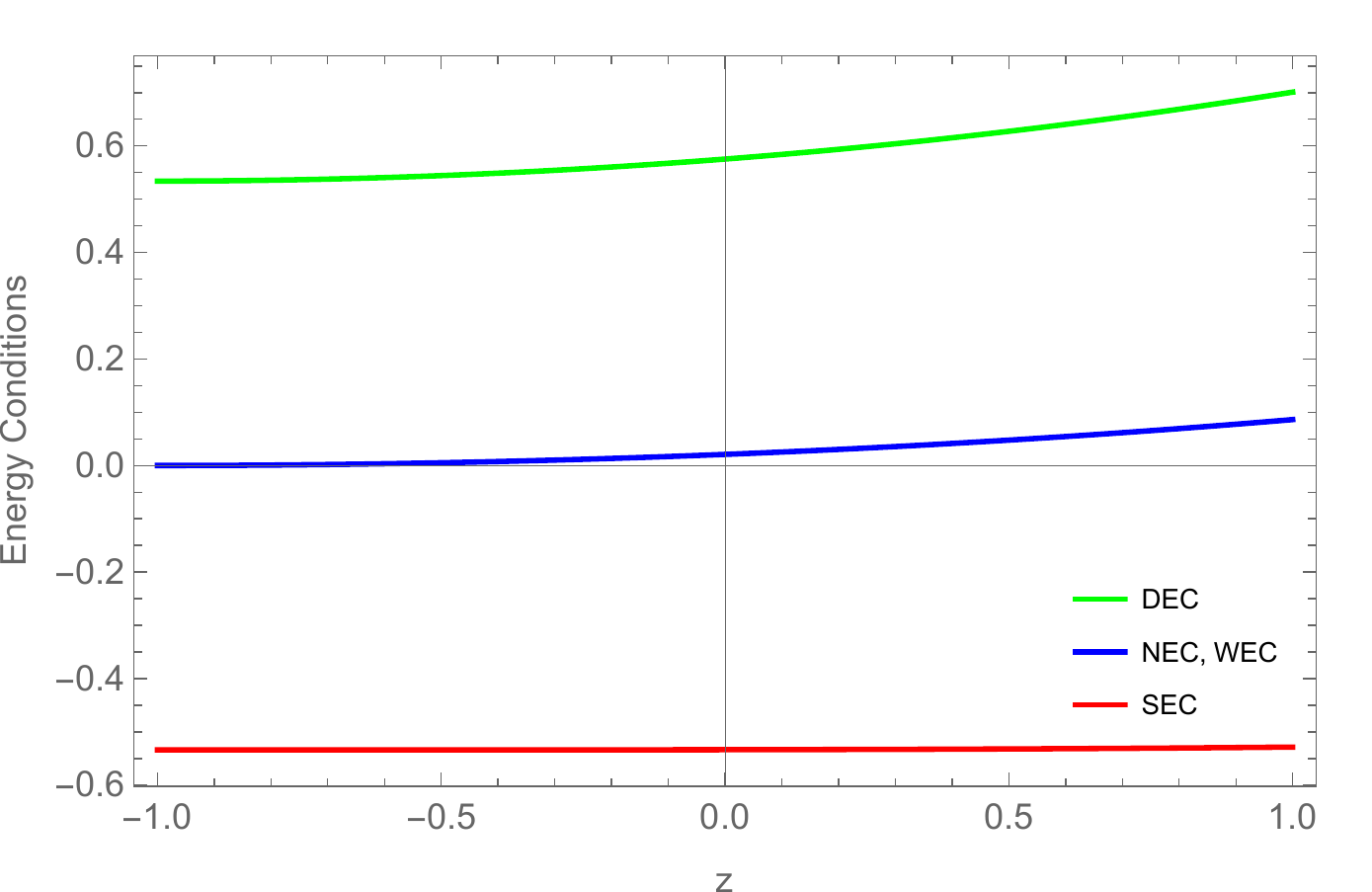}
\includegraphics[width=80mm]{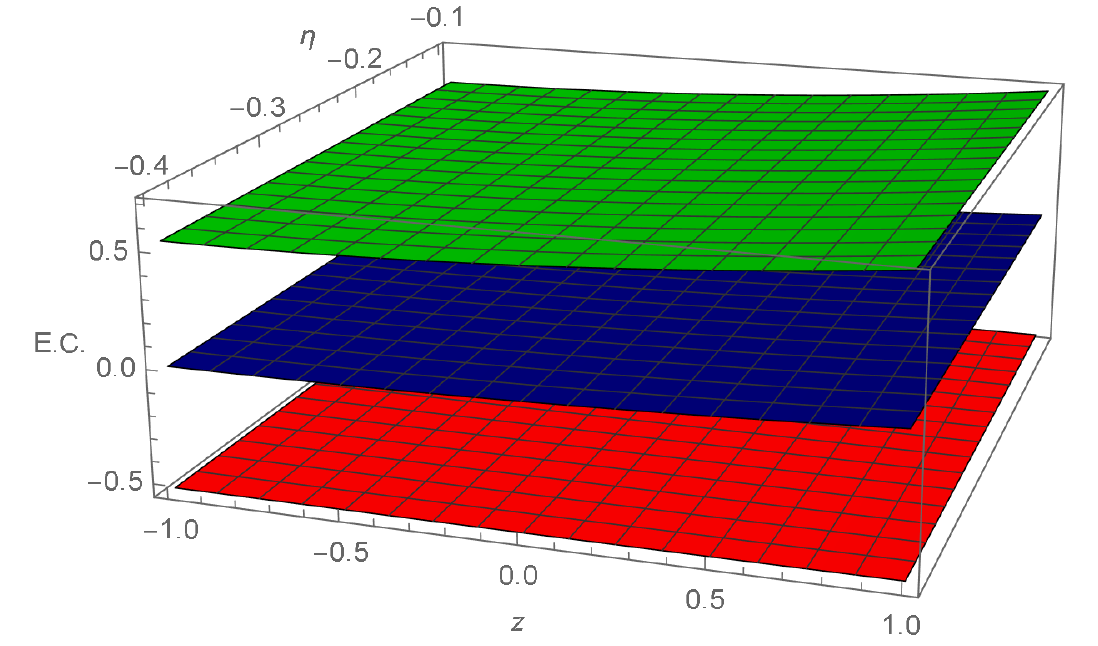}
\includegraphics[width=80mm]{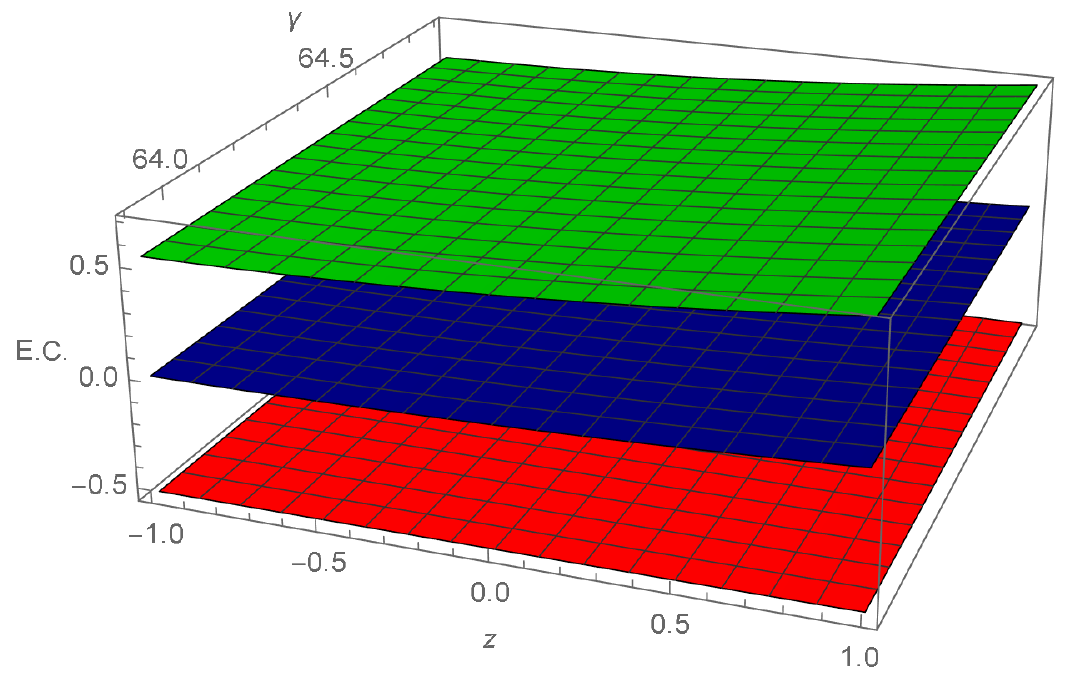}
\caption{Energy conditions in redshift for $\eta=-0.08$, $\gamma=64.9$ (left panel), 3D  plot varying $\eta$ (right panel), 3D  plot varying $\gamma$ (below panel). Other parameters, $\alpha=-0.05$, $T_{0}=105 \pi$.}
\label{Fig3}
\end{figure}

The graphical behaviour of energy conditions are shown in FIG. \ref{Fig3}. The SEC violates throughout the evolution. The NEC remains positive at early time, then merge with the null line and subsequently violates at late time, thereby supports $\Lambda$CDM behaviour. The DEC shows decreasing behaviour and remains positive throughout [FIG. \ref{Fig3}(left panel)]. To get the better visualisation, $3D$ plots with the third dimension as $\eta$ [FIG. \ref{Fig3}(right panel)] and $\gamma$ [FIG. \ref{Fig3}(below panel)] are also presented. The behaviour of energy conditions in all the combinations of the parametric values remain similar. This behaviour further strengthen the claim of the model supporting the late time cosmic acceleration.

\section{Geometric parameters and the scalar perturbations}
\label{sec:Cosmography Parameters}
\begin{figure}[H]
\centering
\includegraphics[width=80mm]{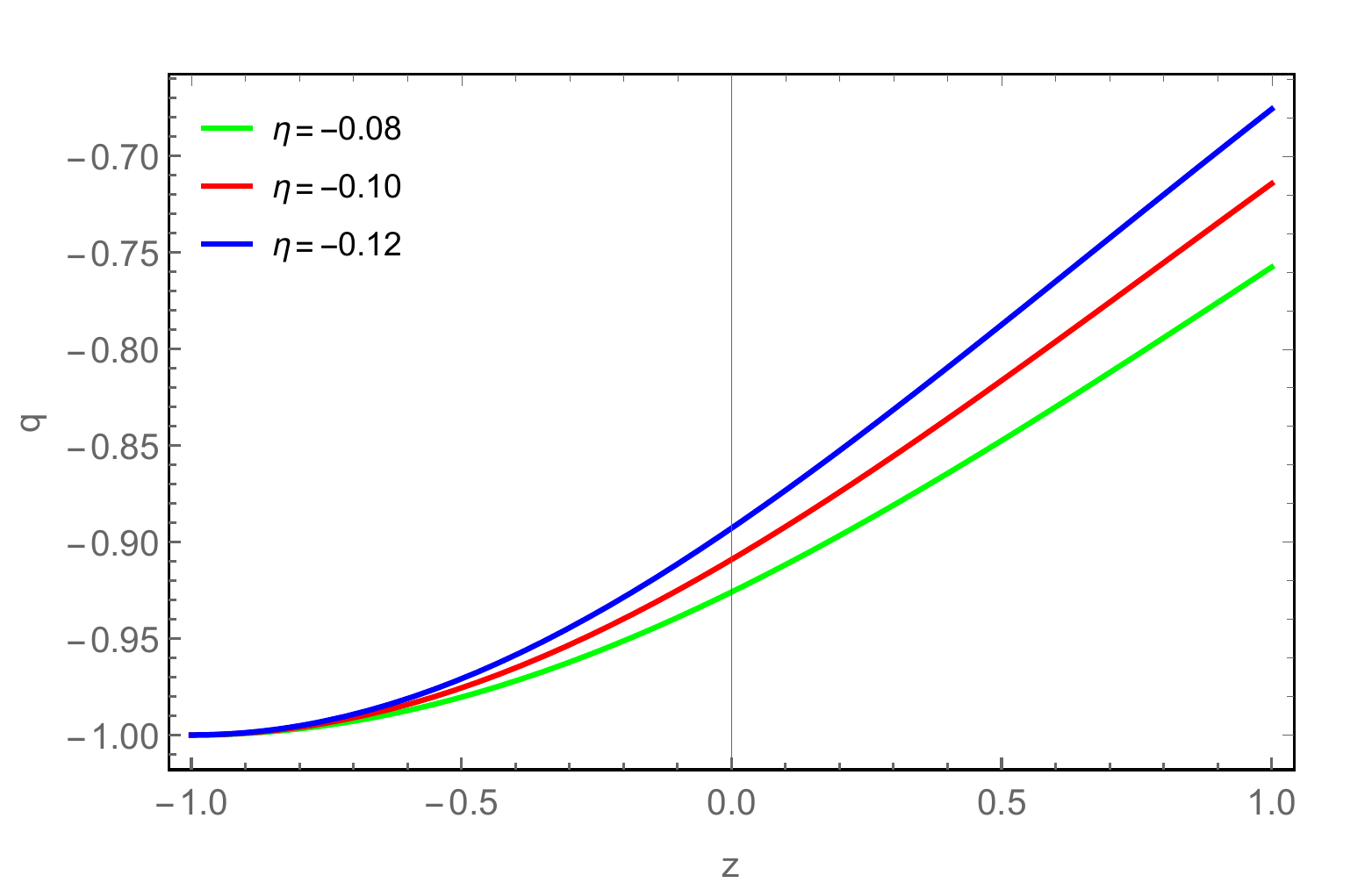}
\includegraphics[width=80mm]{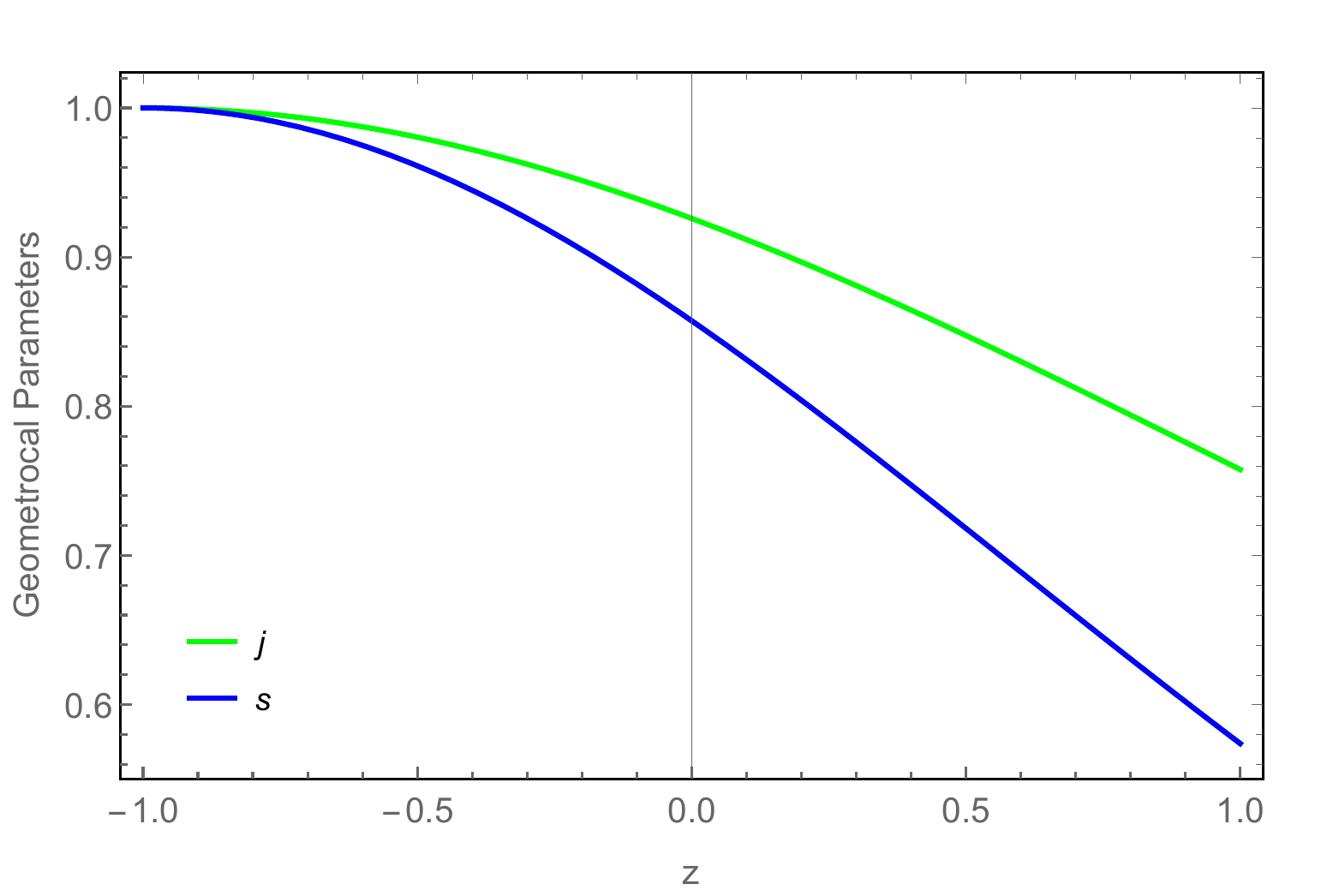}
\caption{Deceleration parameter (left panel) and $(j,s)$ pair (right panel) in redshift.  Other parameters, $\alpha=-0.05$, $T_{0}=105 \pi$.}
\label{Fig4}
\end{figure}

The geometrical or cosmographic parameters have important role in the cosmological model building. The present value of Hubble parameter $H$ and deceleration parameter $q$ required to be aligned with the cosmological observations value for the model to be realistic. Moreover, the decelerating or accelerating behaviour of the model depends on the negative or positive value of $q$. The $(j,s)$ pair respectively denote the jerk and snap parameters that help in distinguishing the dark energy models \cite{Sahni03}. All these parameters are model independent and can be described by expanding the scale factor $a(t)$ in terms of Taylor's series expansion around present time $t_{0}$. In order to lift the model degeneracy, the cosmographic test in the context of $f(T,B)$ gravity has been reviewed in Ref. \cite{Capozziello19}. In the considered scale factor, the geometrical parameters can be obtained in most simplified form as,
\begin{eqnarray} \nonumber
q(z)&=&\frac{1}{\eta  (z+1)^2-1} \,, \\ \nonumber
j(z)&=&-\frac{1}{\eta  (z+1)^2-1} \,,\\ \nonumber
s(z)&=&\frac{1}{\left(\eta  (z+1)^2-1\right)^2} \,.\label{eq:20}
\end{eqnarray} 
The graphical behaviour of the deceleration parameter and $(j,s)$ pair are given in FIG. \ref{Fig4}. The deceleration parameter approaches to $-1$ at late time and at present, $q_0=-0.93,-0.91,-0.89$ respectively for $\eta=-0.08,-0.10,-0.12$. During the entire evolution it remains negative and hence the model shows ever accelerating behaviour [FIG. \ref{Fig4} (left panel)]. We wish to note here that the present value of Hubble parameter obtained as, $H_0=67.42,68.02,68.62$ $km/s/Mpc$ against the Planck 2018 results on cosmological parameters as, $H_0=(67.4\pm 0.5) km/s/Mpc$ \cite{Aghanim20}. The diagnostic pair, $(j,s)$  approach to $1$ at late time, though they start evolving from different phase [FIG. \ref{Fig4}(right panel)]. 

\begin{figure}[H]
\centering
\includegraphics[width=80mm]{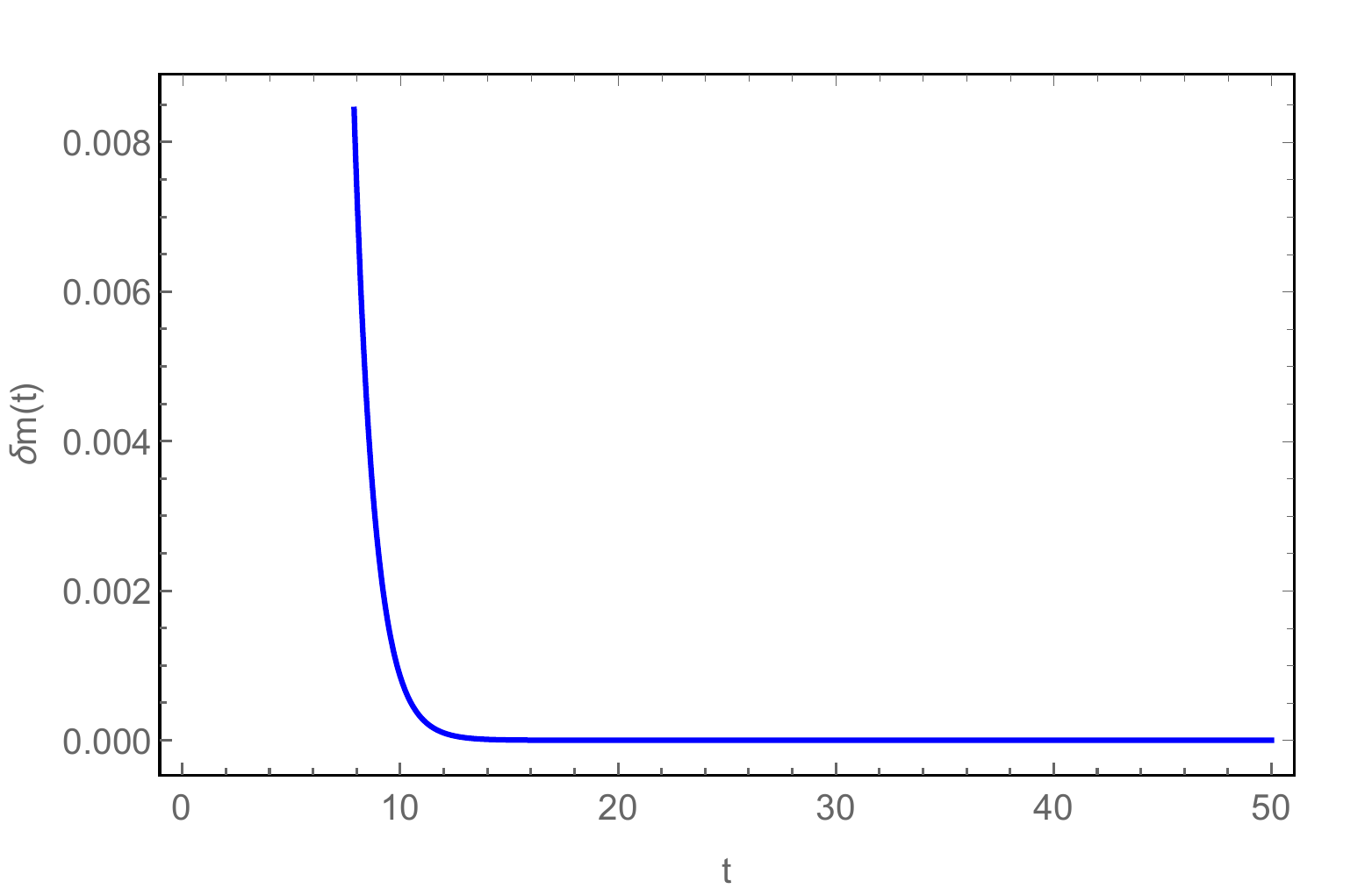}
\includegraphics[width=80mm]{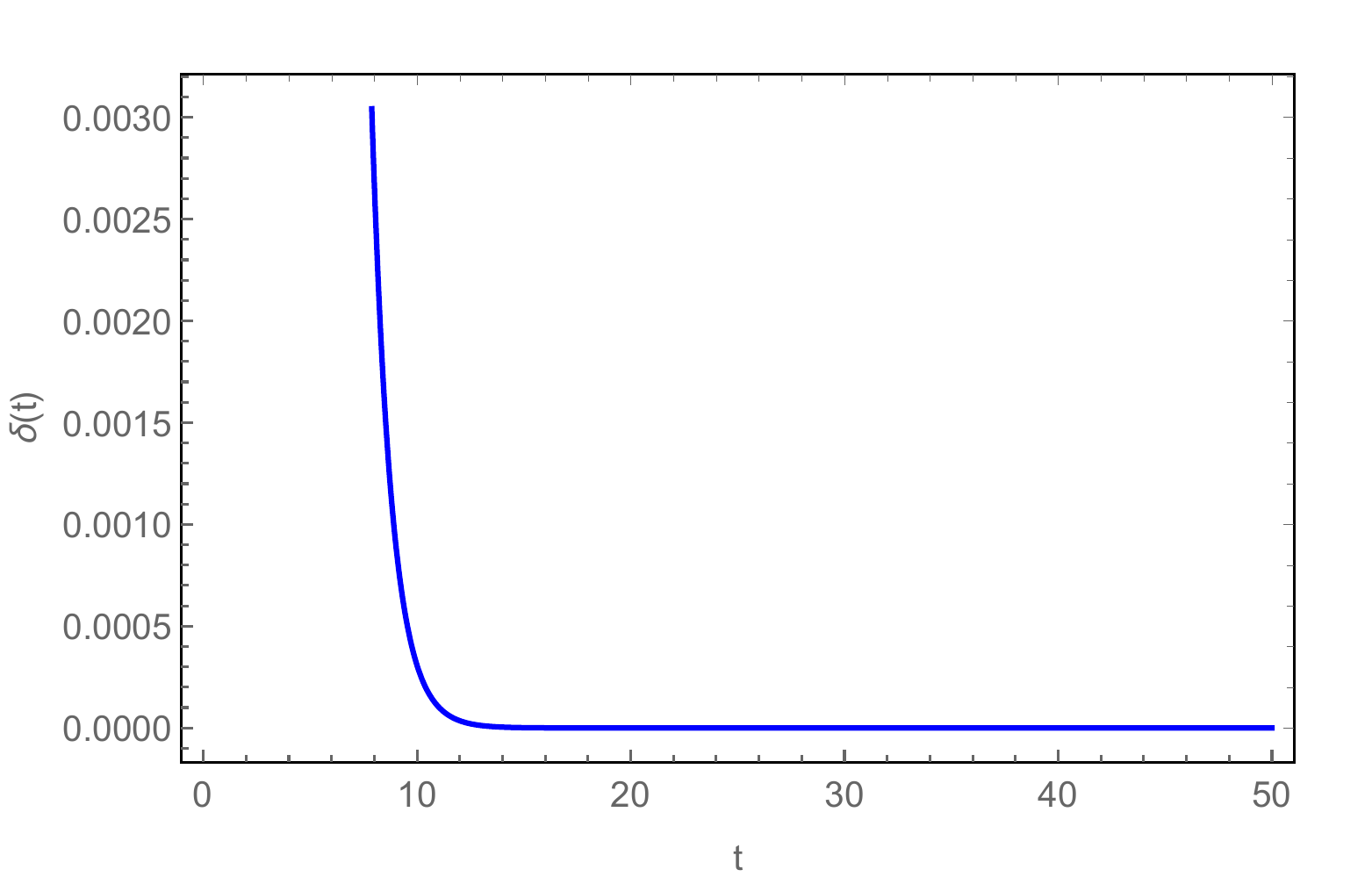}
\caption{Plots of perturbation in the energy density ($\delta_{m}(t)$) (left panel) and Hubble parameter ($\delta (t)$) (right panel) in cosmic time. Other parameter values remain same as before.}
\label{Fig5}
\end{figure}

So far, we have seen that the model shows an accelerating behaviour and at late phase it appears to be in the $\Lambda$CDM phase. In the process, we have considered several assumptions and to understand the viability of these considerations, we shall perform the stability analysis of the model through homogeneous and isotropic linear perturbation. We consider the perturbation of Hubble parameter and the energy density respectively as,
\begin{equation}
H(t)=H_{a}(1+ \delta(t))\,,
\end{equation}  \label{eq:21}
and
\begin{equation}
\rho(t)=\rho_{a}(1+ \delta_{m}(t))\,.
\end{equation} \label{eq:22} 
The functional $f(T,B)=\alpha T log\left(\frac{T}{T_{0}}\right)+\beta B$ can be expanded in powers of $T_{a}$ and $B_{a}$ as\\
\begin{equation}
f(T,B)=f_{a}+\left(\alpha log\left[\frac{T_{a}}{T_{0}}+\alpha\right]\right)(T-T_{a})+\beta(B-B_{a})+\mathcal{O}^{2}\,.
\end{equation} \label{eq:23}
Where $\mathcal{O}^{2}$ includes second order and higher order derivatives of $T$ and $B$ and the subscript $a$ denotes the functional $f(T,B)$ and its derivatives evaluated at $T=T_{a}$ and $B=B_{a}$. We shall consider the linear terms of the defined perturbation, hence the FLRW space-time in perturbative approach resulted in
\begin{equation}\label{eq:24}
-6H_{a}^{2} \beta  \dot{\delta}(t)-(48 H_{a}^{2}\beta+12 \dot{H}_{a} \alpha+24 H_{a}^{2}\alpha)\delta(t)=\kappa^{2} \rho_{a} \delta_{m}.
\end{equation}
Now from the energy density expression we get,
\begin{equation}\label{eq:25}
\rho_{a}=-3H_{a}^{2} \alpha log\left[\frac{T_{a}}{T_{0}}\right]-T_{a} \alpha\,,
\end{equation}
The conservation expression represents the relationship between $\delta_{m}(t)$ and the Hubble parameter which may be written as,
\begin{equation}\label{eq:26}
\dot{\delta}_{m}(t)+3H_{a}(t)\delta(t)=0\,,
\end{equation}
Here on referring Eq. \eqref{eq:25} and on neglecting higher derivative term of $\delta (t)$ in Eq. \eqref{eq:24}, the relationship between Hubble parameter at $H_a$ and $\delta(t)$ can be obtained as,
\begin{equation}\label{eq:27}
-(48 H_{a}^{2}\beta+12 \dot{H}_{a} \alpha+24 H_{a}^{2}\alpha)\dot{\delta}(t)=\kappa^{2}18H_{a}^{3}\alpha\delta(t)\,, 
\end{equation}
On solving, we obtain the perturbation parameters $\delta(t)$ and $\delta_{m}(t)$ as, 
\begin{equation}
\delta(t)=\tau  \exp \bigg\{\frac{3 \alpha  \left(\beta  \log \left(2 \alpha +3 \beta -2 (\alpha +2 \beta ) \cosh ^2(\gamma  t)\right)-4 (\alpha +2 \beta ) \log (\cosh (\gamma  t))\right)}{8 (\alpha +2 \beta ) (2 \alpha +3 \beta )}\bigg\}\,,
\end{equation}\label{eq:28}

\begin{equation}
\delta_{m}(t)=\frac{\delta(t)}{6 \alpha  \gamma ^3}\bigg\{ \frac{\coth ^3(\gamma  t) \left(24 \alpha  \gamma ^2 \tanh ^2(\gamma  t)+48 \beta  \gamma ^2 \tanh ^2(\gamma  t)+12 \beta  \gamma ^2 \text{sech}^2(\gamma  t)\right)}{6 \alpha  \gamma ^3}\bigg\}\,.
\end{equation}\label{eq:29}
where $\tau$ is an integrating constant. The graphical representation for perturbation in the energy density ($\delta_{m}(t)$) and Hubble parameter ($\delta(t)$) is presented in Fig. \ref{Fig5}. Both the plots are lying in the positive region, and showing decreasing behaviour as time goes on increasing. Both the plots, $\delta_{m} (t)$ and $\delta(t)$ approach to zero for increasing cosmic time thereby ensuring the stability of the model. 

\section{Results and Conclusion}\label{sec:conclusion}

We have presented an accelerating cosmological model that shows the $\Lambda$CDM behaviour at late times of the evolution. Initially we have derived the dynamical parameters with some assumed form of the function $f(T,B)$ and to understand the cosmic evolution, we adhered to the hyperbolic function of the scale factor. The geometrical parameters are scale factor dependent and the value of the scale factor parameters are significant to frame a realistic model. Accordingly, the scale factor parameters are adjusted to obtain the value of the geometrical parameters as supported by the cosmological observations. In this model the present value of deceleration parameter and Hubble parameter are obtained in the range $[-0.93, -0.89]$ and $[67.42, 68.62]$ respectively. To understand the dynamics of the model, the EoS parameter has been analysed and with the already adjusted value of the scale factor and model parameters, the present value of EoS parameter has been obtained. Though at the late time of the evolution it shows the $\Lambda$CDM behaviour, at present time its values are noted in the range $[-0.942,-0.916]$. The present value of geometrical and EoS parameter obtained in the model are in accordance with recetn cosmological observations and the same has been discussed in the respective section. One important note is that while deriving the dynamical parameters, the terms containing $\beta$ identically vanish. Therefore the role of boundary term in the evolution process could not be assessed. However, since it depends on the Hubble parameter, it has been observed that it remains positive throughout and reduces over time. \\

Another important aspect of adjusting the parameters is to keep the energy density positive throughout and here we obtained the same with a gradually decreasing energy density. The energy conditions are studied to check the viability of the model. Violation of SEC supports the accelerating behaviour. The NEC gets validated at an early time and vanishes at a late time. The DEC is validating throughout the evolution with decreasing behaviour. In this model, the violation of SEC in entirety and of NEC at late time have been shown. Since assumptions are made to obtain the expressions of the parameter, the stability of the model needs to be checked. We have used the linear perturbation approach to study the stability of the model. As the energy density and Hubble parameter perturbation shows decreasing behaviour and approach to zero for increasing values of cosmic time $t$, the stability of the model has been confirmed. So, in conclusion, we can mention that in the context of teleparallel gravity with the boundary term, the accelerating behaviour of the Universe can be realised. 

\section*{Acknowledgement}
SAK acknowledges the financial support provided by University Grants Commission (UGC) through Junior Research Fellowship (UGC Ref. No. : 191620205335) to carry out the research work. BM and SKT acknowledge IUCAA, Pune, India for hospitality and support during an academic visit where a part of this work has been accomplished. The authors are thankful to the honorable referees for their comments and suggestions for the improvement of the paper.

\end{document}